\documentclass{emulateapj}

\newcommand \thj {\theta}

\newcommand \toff {t_{\rm off}}
\newcommand \roff {r_{\rm off}}
\newcommand \tem {\tilde t}
\newcommand \temin {\tilde t_{\rm init}}
\newcommand \temoff {\tilde t_{\rm off}}
\newcommand \tcol {\tilde t_{\rm col}}
\newcommand \rcol {r_{\rm col}}
\newcommand \rin {r_{\rm init}}
\newcommand \tho {\tilde\theta}
\newcommand \thobs {\theta_{\rm obs}}
\newcommand \thc {\theta_{\rm c}}

\newcommand \nuobs {\nu_{\rm obs}}
\newcommand \gamc {\Gamma_{0}}

\newcommand \dapp {D_{\rm app}}
\newcommand \gsm  {\Gamma_{s}}
\newcommand \gla  {\Gamma_{l}}
\newcommand \bsm  {\beta_{s}}
\newcommand \bla  {\beta_{l}}
\newcommand \dtsp {\Delta t_{\rm spt}}
\newcommand \dtsph {\Delta t_{\rm sph}}
\newcommand \dttem {\Delta t_{t}}

\newcommand \beq {\begin{equation}}
\newcommand \eeq {\end{equation}}

\def\la{\hbox{\hspace{1.5mm}}\raise2pt
       \vbox{\hbox{$<$}}\lower2pt
       \vbox{\moveleft9.0pt\hbox{$\sim$ }}\hbox{\hskip 0.05mm}}
\def\ga{\hbox{\hspace{1.5mm}}\raise2pt
       \vbox{\hbox{$>$}}\lower2pt
       \vbox{\moveleft9.0pt\hbox{$\sim$ }}\hbox{\hskip 0.05mm}}
\def\npar{\hbox{\hspace{1.0mm}}
       \vbox{\hbox{$\parallel$}}
       \vbox{\moveleft6.3pt\hbox{$\not$}}\hbox{\hskip 0.5mm}}

\shorttitle{Curvature effect in structured GRB jets}
\shortauthors{Dyks, Zhang \& Fan}

\begin{document}

\title{Curvature effect in structured GRB jets}

\author{J. Dyks$^{1,2}$, Bing Zhang$^{1}$, and Y. Z. Fan$^{1,3,4}$}
\affil{$1$ Physics Department, University of Nevada Las Vegas, NV, USA
\\
$2$ Centrum Astronomiczne im.~M.~Kopernika PAN, Toru{\'n},
Poland \\
$3$ Purple Mountain Observatory, Chinese Academy of Sciences,
Nanjing 210008, China \\
$4$ National Astronomical Observatories, Chinese Academy of
Sciences, Beijing, 100012, China}
\email{jinx@ncac.torun.pl}%, bzhang@physics.unlv.edu, yzfan@pmo.ac.cn}

\begin{abstract}
We study the influence of jet structure on the curvature effect
in a GRB-lightcurve. 
Using a simple model of jet emissivity, we 
numerically calculate
lightcurves for a short flash from a relativistic outflow
having various profiles of the Lorentz factor
and outflowing energy density (gaussian, core+power-law).
We find that for ``on-beam" viewing
geometry, with the line of sight passing through the bright core of
the outflow, the shape of the lightcurve practically does not depend
on the jet structure, initially following the temporal slope
$2+\delta$, where $\delta$ is the spectral index. When the viewing
angle is larger than the core, the light curve decaying slope is
shallower. We discuss the implications of our results for the {\em
Swift} data.
\end{abstract}


\section{Introduction}

Recent {\em Swift} observations of gamma-ray burst afterglows
(Tagliaferri et al.~2005; Chincarini et al.~2005; Nousek et al. 2005;
O'Brien et al. 2005) revealed very fast drops of X-ray flux
a few hundred seconds after the trigger time. The rapid decay is
usually also followed with the X-ray flares (Burrows et al. 2005;
O'Brien et al. 2005).
A possible mechanism of this phenomenon is a sudden decrease
of jet's emissivity (see Zhang et al.~2005 for more comprehensive
discussion of possible scenarios). For a curved front surface of a jet
the turn-off
is perceived by an observer as a fast, yet not abrupt, 
decrease of flux, because
the simultaneous drop of emissivity at different angles $\tho$
from the line of sight is perceived at different moments.
For uniform properties of the jet, and in the absence of further 
complicating factors, the temporal slope $\alpha$ of a lightcurve 
resulting from such a ``curvature effect" is equal to $2+\delta$, 
where $\delta$
is the spectral index and the conventions 
$F_\nu \propto t^{-\alpha}$ and $F_\nu \propto \nu^{-\delta}$
are assumed
(Kumar \& Panaitescu 2000; Fan \& Wei 2005; Zhang et al. 2005;
Panaitescu et al.~2005). 

In addition to the spectral properties of emitted radiation,
the structure of the jet, i.e.~the variation of jet properties
as a function of the angle $\thj$ from the jet axis, may influence
the observed lightcurve -- an issue addressed in many recent investigations
(e.g. Zhang \& M\'esz\'aros 2002; Rossi et al. 2002; Wei \& Jin 2003;
Kumar \& Granot 2003; Granot \& Kumar 2003; Salmonson 2003; Zhang et
al. 2004; Yamazaki et al.~2005).
During the curvature effect, an observer receives radiation 
emitted from different regions of the outflow (different $\thj$). 
Therefore, an interesting question is whether the curvature effect
depends on the unknown jet structure.
The main purpose of this Letter is to address this question.

\section{Calculation method}
\label{cmethod}

To calculate the lightcurves we use a three dimensional
code which rigorously takes into account all kinematic effects
that affect the observed flux (eg.~Doppler boost, propagation time delays)
in the way described by Salmonson (2003).
To minimize the number of parameters involved,
the curvature effect is studied for a short flash, with emission
lasting between $\temin$ and $\temoff=1.05\temin$ in the frame of the central
source (we take $\temin=10^{14}\ {\rm cm}/c$, where c is the speed 
of light). This is the typical internal shock radius (Rees \&
M\'esz\'aros 1994), at which the dynamics of the outflow evolves
negligibly.   
At the time $\temin$ the outflow is located at the radial distance
$\rin = \beta(\thj)\temin$, where $\vec\beta=\vec v/c$ is the velocity
of the outflow at the angle $\thj$ from the jet axis, and $\beta =
|\vec \beta|$. This corresponds to the choice of $\tem=0$
for the moment of ejection (see Zhang et al.~2005 for the discussion
of the ``zero time effect"), and implies that the surface of the structured
jet is slightly non-spherical during the flash.\footnote{Complications
introduced by the $\thj$-dependent
Lorentz factor are discussed in Section \ref{location}.}

For the emissivity in the comoving frame we assume:
\begin{eqnarray}
I^\prime_\nu(\thj) \propto (\nuobs D)^{-\delta}\ \frac{\epsilon}{\Gamma}\ \ \ \ {\rm if}
\ \ \Gamma(\thj) \ge 10
\label{int}\\
I^\prime_\nu(\thj) = 0\ \ \ \ {\rm if}\ \ \Gamma(\thj) < 10
\label{gmin}
\end{eqnarray}
where $\nuobs$ is the observed frequency, 
$D=\Gamma^{-1}(1 - \vec \beta\cdot \hat n)^{-1}$ 
is the Doppler 
factor, $\hat n$ is a unit vector pointing toward the observer,
$\epsilon$ is the energy of the outflow per unit solid angle, 
and
$\Gamma=(1-\beta^2)^{-1/2}$
is the bulk Lorentz factor of the outflow.
The quantities $\epsilon$ and $\Gamma$ in eq.~(\ref{int})
(and thereby $\beta$ and $D$), all refer to the angle $\thj$
from the jet axis and the redshift $z=0$ is assumed.
The factor $\epsilon/\Gamma$ is introduced to represent the 
number of electrons at different viewing angles,
i.e. $N_e(\thj)\propto \epsilon/\Gamma$.

The code is able to deal with any axially symmetric jet structure.
After testing the code on the ``top hat" jet case, we have made 
calculations for four jet structures:
\begin{enumerate}
\item
gaussian, with
\begin{equation}
\epsilon(\thj) \propto \exp\left(-\frac{1}{2}
                                 \frac{\thj^2}{\thc^2}\right),
				 \ \ \ \ 
\Gamma(\thj) = 1 + (\gamc - 1)\exp\left(-\frac{1}{2}
       \frac{\thj^2}{\thc^2}\right),
\label{gau}
\end{equation}
where $\gamc=\Gamma(\thj=0)$ and $\thc$ is the characteristic
width of the jet;
\item
``e-gaussian" with $\epsilon(\thj)$ given by eq.~(\ref{gau})
but with a fixed $\Gamma=\gamc$ for all angles;
\item
core $+$ power law (hereafter CPL case):
\beq
\epsilon(\thj) \propto \left(1 + \frac{\thj^2}{\thc^2}\right)^{-a/2},
    \ \ \ \ 
\Gamma(\thj) = 1 + (\gamc - 1)
    \left(1 + \frac{\thj^2}{\thc^2}\right)^{-b/2},
\label{cpleq}
\eeq
with $a=2$, $b=2$;
\item
``e-CPL" case, which differs from CPL only in that $\Gamma$ 
is fixed ($b=0$ in eq.~\ref{cpleq}).
\end{enumerate}

Thus, we consider two qualitatively different scenarios:
cases 1 and 3 with both $\Gamma$ and $\epsilon$ decreasing towards
the edge of the jet according to (nearly) the same law, and cases 2 and 4
with $\epsilon$ decreasing, but $\Gamma$ fixed.
Note that in cases 1 and 3 the emissivity in the comoving frame is
uniform since $\epsilon/\Gamma \approx$ constant, as is in the top-hat
case. The variations of $\Gamma$ influence  
eq.~(\ref{int}) only through the Doppler factor D.
Our calculation in the cases 1 and 3 
is limited only to the relativistic part of the outflow
(eq.~\ref{gmin}).

\section{Results}

Figs.~\ref{f1}a to \ref{f1}d present our results obtained for
$\gamc = 300$, $\thc=5.7^\circ$, and $\delta = 1.3$.
Different lightcurves, shown in the top panels, correspond
to different viewing angles $\thobs$. 
The solid and dotted lines correspond to the ``on-beam" viewing
($\thobs=0$ and $0.5\thc$, respectively). The short dashed line is for
viewing along the edge of the bright jet core ($\thobs = \thc$), whereas
the other lines are for ``off-beam" viewing ($\thobs > \thc$).

\subsection{The ubiquitous slope $2+\delta$}

\begin{figure}
\epsscale{1.15}
\plotone{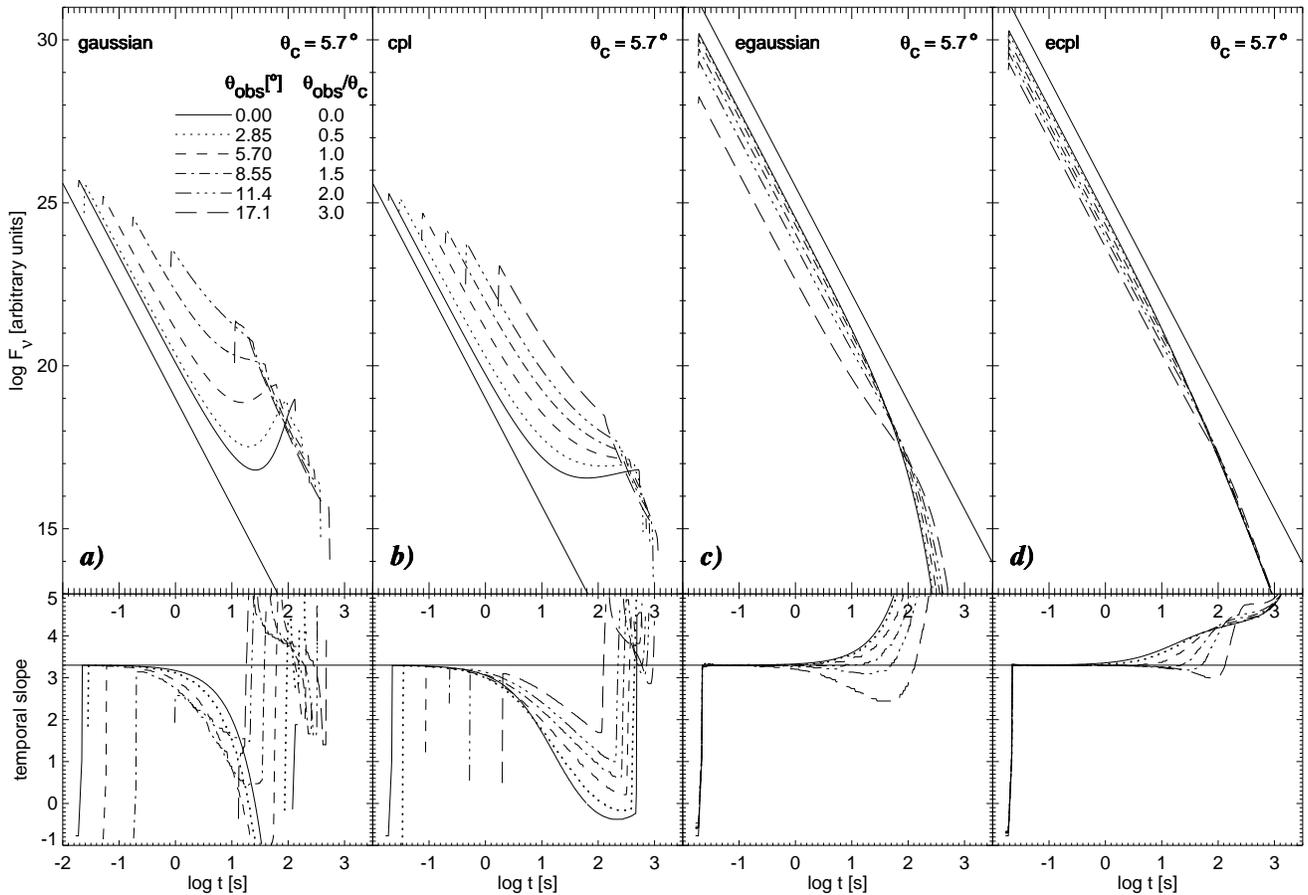}
\caption{{\it Top:} Lightcurves for a short flash from a structured jet
illustrating the curvature effect. 
Columns a), b), c) and d) refer to the structures described 
as cases 1, 2, 3, and 4 in Section \ref{cmethod}. 
Different lines: 
solid, dotted, dashed, dot-dashed, 
dot-dot-dot-dashed, and long-dashed, correspond to the viewing angles
$\thobs/\thc=0$, $0.5$, $1$, $1.5$, $2$, and $3$, respectively.
The straight solid line is for the analytically predicted slope
$2+\delta$. The result is for $\delta=1.3$, $\thc = 5.7^\circ$,
and $\gamc = 300$.
{\it Bottom:} Temporal slope for the lightcurves in the top panel.
\label{f1}}
\end{figure}

The straight solid line in top panels marks
the slope $2+\delta=3.3$, analytically predicted for a uniform
outflow. 
It can be seen that regardless of the jet structure,
majority of modeled lightcurves
follows the $2+\delta$ slope for many orders of magnitude in flux.
The only exceptions are the lightcurves for off-beam viewing
of a jet with $\Gamma$ decreasing with distance from the jet axis
(gaussian case, Fig.~\ref{f1}a, and CPL case, Fig.~\ref{f1}b).
These off-beam lightcurves exhibit a break at a late time 
($t \ga 30$ s for the gaussian case, Fig.~\ref{f1}a, and 
$t \ga 10^2$ s for the CPL case, Fig.~\ref{f1}b),
after which the flux drops rapidly (on average following the slope $\sim 3.8$).
The fast drop is caused by the lack of emission from the outer parts
of the outflow (the jet edge given by eq.~\ref{gmin} starts
to be visible for an observer).

\begin{figure}
\epsscale{0.5}
\plotone{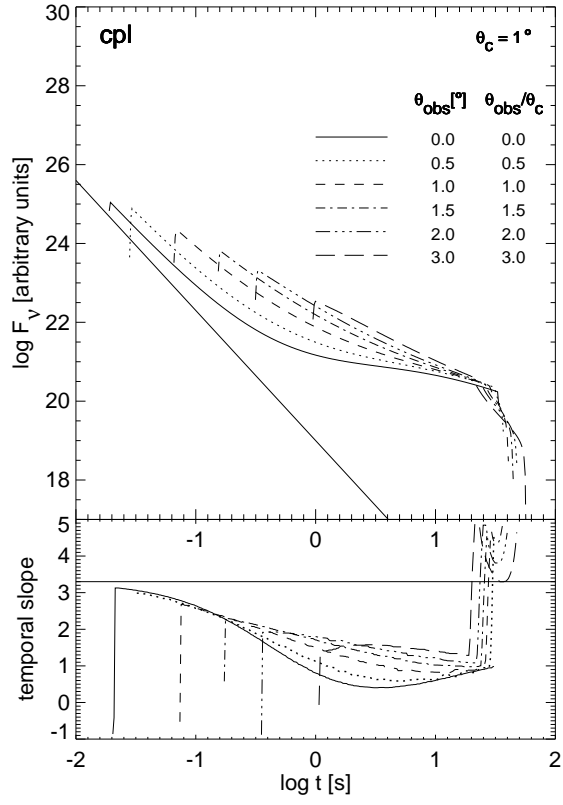}
\caption{Same as in Fig.~\ref{f1}b (core $+$ power law case)
but for a significantly smaller core width $\thc = 1^\circ$.
\label{f2}}
\end{figure}

The main reason for the trend to initially follow the $2+\delta$ slope
is that the decrease of flux due to the curvature
effect occurs on a much shorter timescale than that for
the jet structure to take effect.
For the spectral index $\delta$ the flux decreases by $m$ orders
of magnitude after a time of $t_{\rm crv}=10^{m/(2+\delta)}\toff$,
where $\toff$
is the observer time at which the curvature effect began.
For a typical $\delta \sim 1$, the flux drops
by one-order of magnitude after a short time
$t_{\rm crv}\sim 2\toff$. A drop of three
orders of magnitude occurs in no more than a decade in time.
On the other hand, for a spherical surface of the jet,\footnote{
The front-surface of a jet with $\thj$-dependent Lorentz factor
should be non-spherical
because the outflow velocity is not uniform. 
In the cases shown in Figs.~\ref{f1} and \ref{f2}, this effect is
small and can be neglected (see Section \ref{location}).
\label{futadruga}
}
the observer can perceive the switch-off of emissivity
at angle $\tho$ measured from the line of sight at a time
$t_{\rm \tho} \simeq [1 + (\tho\Gamma)^2]\toff$.
One can see that the structure of the outflow must have
the typical angular scale of the order of $3/\Gamma$ to affect the observed
flux before $10\toff$. For $\gamc > 10^2$, the parameters of the outflow
would have to vary strongly on a scale of one degree
to be detectable at early time, when the flux is high.
In both cases considered in this paper (i.e.~gaussian and CPL),
$\Gamma$ changes little within the core
of a jet, and this is why $\alpha \simeq 2+\delta$ for $\thobs < \thc$.
The strong deviation from $2+\delta$ starts to be pronounced only when
$\thobs \sim \thc \gg 1/\gamc$, i.e.~at late times or large viewing angles.

For a small core of angular radius $\thc = 1^\circ$
the discrepancy between the calculated slope
and $2+\delta$ appears at earlier time, as can be seen in
Fig.~\ref{f2}.

\subsection{Rebrightening}

At late observational times the lightcurves diverge from the $2+\delta$ slope,
even in the ``on beam" case.
An interesting feature that appears for outflows with 
$\thj$-dependent Lorentz factor is a rebrightening,
visible for the gaussian case (Fig.~\ref{f1}a) at $t > 30$ s.

The reason for this behaviour is a nonmonotonic dependence of
Doppler factor $D$ on $\thj$. 
After the sudden turn-off of emissivity, the observed flux
decreases as $D^{(2+\delta)}$, and 
the Doppler factor can be approximated with
$\dapp = 2 [\Gamma(\thj)]^{-1}\thj^{-2}$ in the limit of $\thj \gg
1/\Gamma$. Initially, the Lorentz factor changes slowly ($\dapp
\approx 2 /(\gamc \thj^2)$),
so that $\dapp$ decreases with increasing $\thj$ (and $t$),
and the flux follows the slope $2+\delta$.
As soon as $\thj(t) \sim \thc$, the factor $\Gamma(\thj)$ starts to decrease
fast which initially slows down the rate of the flux decrease
and the discrepancy from the slope $2+\delta$ appears. 
Eventually, the decrease of $\Gamma(\thj)$
overcomes the dependence $\dapp \propto \thj^{-2}$
and causes $\dapp$ (as well as the observed flux) to increase with time.

In the gaussian case (Fig.~\ref{f1}a) the flux reaches minimum
when the observer receives radiation from $\thj = 2^{1/2}\thc$,
which happens near $t \simeq 30$ s in Fig.~\ref{f1}a.
The Lorentz factor at this angle is $\Gamma = 1+(\gamc-1)\exp(-1)
\approx 111$ (for $\gamc = 300$ used throughout this paper).

In the CPL case shown in Fig.~\ref{f1}b the flux at $t \ga 10^2$ s 
is nearly constant, because of the onset of the regime
$\thj \gg \thc$, in which the specific structure of the outflow
($\Gamma(\thj) \propto \thj^{-2}$)
cancels out the $\thj$-dependence in 
$\dapp \propto \Gamma^{-1}\thj^{-2}$.
For the CPL case with a smaller core (Fig.~\ref{f2}) the levelling
happens at earlier time ($t \sim 1$ s).
After the plateau phase the lightcurve tends
to steepen again\footnote{
The asymptotic slope which the lightcurve is trying to take on
(but never succeedes because of the break)
is equal to $\alpha=[\delta + 4 - (2/n)]/2$, where
$n \ge 1$ is the exponent in the relation $\Gamma(\thj) \propto \thj^{-n}$
($n=2$ in the CPL case)
and the convention $F_\nu \propto t^{-\alpha}$ was assumed.
This temporal index corresponds to the limit $\thj \ll 1/\Gamma$.}
before undergoing the jet edge break.

\subsection{Role of the emission preceding the turn-off}

The lightcurves with the rebrightening effect can be considered
as consisting of two 
qualitatively different parts: one of them has a power-law shape
with a slope very close to $2+\delta$; the other part,
which includes the rebrightening, has a variable slope.
This specific shape provides a good insight into the problem of how
the shape of a lightcurve depends on the emissivity which preceded 
the turn-off.

\begin{figure}
\epsscale{0.5}
\plotone{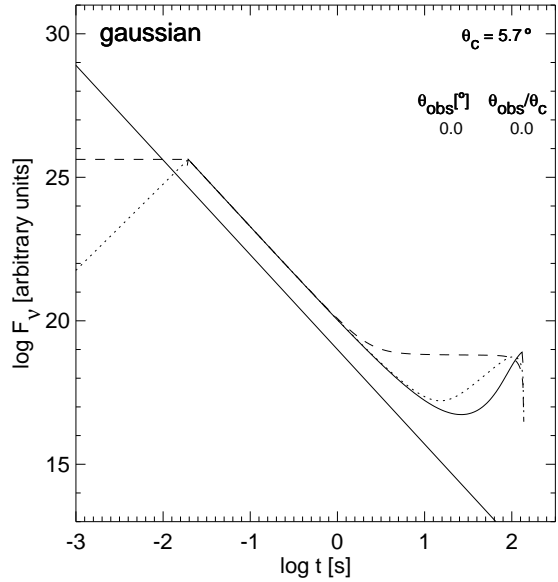}
\caption{Comparison of a ``short flash" lightcurve (solid curve)
with lightcurves for the ``sudden turn-off" case.
The latter differ in a radial dependence of emissivity before the turn-off:
the dotted line
is for $I^\prime_\nu \propto r^3$, and
the dashed one is for $I^\prime_\nu = const$. Note that
the power-law part of the lightcurve after $\toff \approx 0.02$ s
is not affected by the emission
preceding the turn-off.
\label{preced}}
\end{figure}

We have numerically tested a variety of cases, assuming different duration of
emission before the turn-off, and various radial dependence of the emissivity.
We have found that the flux initially drops following the familiar
${2+\delta}$ power-law, \emph{independently on the emissivity that
preceded the turn off.} 
However, the shape of a lightcurve in the region of the variable slope
(especially in the valley preceding the rebrightening)
is very sensitive to the duration, and to the time dependence of emissivity
which preceded the turn-off (Fig.~\ref{preced}).

To understand this result, it is useful to consider the continuous emission
preceding the turn-off, as consisting of a series of short flashes,
each of which contributes its own lightcurve with the shape described
above (power-law $+$ rebrightening) but shifted horizontally with respect
to each other and differently normalized. 
The observed lightcurve can then be considered to be a \emph{sum} of these 
sub-lightcurves.
In the period of time within which all these sub-lightcurves have
the power-law shape (with the slope $2+\delta$),
the total lightcurve will have the same shape.
In the region of the variable slope (near the rebrightening), 
the observed lightcurve
takes on a different shape than the shape of individual components.
Analogous effect happens when one calculates a photon spectrum 
by integrating
over an electron energy distribution having a broken power-law shape.

Therefore,
in the region of the non-power-law shape the lightcurves 
calculated for the short flash
may differ noticeably from those calculated
for the ``sudden turn-off" case. Our choice of the ``short flash"
scenario has been mainly dictated by two facts: 
1) The differences between the two cases in the non-power-law region
are small if the emissivity preceding the turn-off
increases quickly with time, or has a short duration. According to the
internal shock model of GRBs (Rees \& M\'esz\'aros 1994) both these
conditions are likely to be fulfilled in reality.
2) A considerably larger number of model parameters
is required to model the sudden switch-off case (e.g.~the duration
and the radial dependence of emissivity before the turn-off).

\subsection{More realistic models}
\label{location}

For an outflow with the $\thj$-dependent Lorentz factor, the standard
version of a 
curvature effect with the turn-off/flash occuring simultaneously
at the same radial distance $\roff$ from the central object is
hardly conceivable, because of the non-uniform velocity of the outflow.

Therefore, we have also calculated lightcurves for
a locally-short brightening
of a spherical surface. The emission
given by eq.~(\ref{int}) was limited to the region 
between $\rin=R_0$ and $1.05R_0$
and started non-simultaneously at a $\thj$-dependent time
$\temin = r/(\beta(\thj)c)$ (which may roughly correspond to a collision
of a non-uniform outflow with a spherical layer of material).

Lightcurves calculated for such ``non-simultaneous flash" at a fixed $r$ 
are practically identical to those that we get for the
simultaneous flash at the $\thj$-dependent distance. 
This is because the additional radial delay 
$\Delta t_{\rm str} = (r/c)[1 - v(\thj)/v(\thj=0)]$ 
associated with the extra curvature is much smaller
than the standard delay for the spherical curvature: 
$\Delta t_{\rm sph} = (r/c)[1 - \cos\thj]$.
In other words, the reason is the proximity of $\beta(\thj)$ to $1$
within the relativistic part of the outflow.

We next proceed to discuss the case of a flash at a place and time
that correspond 
to a collision of two structured shells: a slow one, characterized by
the Lorentz factor profile $\gsm(\thj)$, and a fast one [$\gla(\thj)$]
ejected from the central source $\delta t$ seconds later.
For $\tem = 0$ corresponding to the ejection of the fast shell,
the collision occurs at the time 
$\tcol(\thj) = \delta t / (\bla(\thj)/\bsm(\thj) - 1)
\approx 2\delta t [\gsm(\thj)]^2$
and at the radial distance $\rcol(\thj) = \bla(\thj)c \tcol(\thj)\approx
2c\delta t [\gsm(\thj)]^2$.
Therefore, for outflows with $\Gamma$
quickly decreasing with $\thj$, as in the cases 1 and 3,
the emission region is very elongated and has a pencil-like shape
($\rcol \propto \gsm^2$).
Even at a large $\thj$ the radiation is emitted from 
$\rcol(\thj) \ll \rcol(\thj=0)$, i.e.~from regions located
very close to the jet axis.

Surprisingly, however, the large additional distance that must be
covered by photons does not produce large discrepancy from
the $2+\delta$ slope.
The reason is that the collision at large $\thj$ occurs at 
much earlier time ($\tcol \propto \gsm^2$).
This temporal advance [$\dttem = \tcol(0) - \tcol(\thj)$;
hereafter $(0)$ means ($\thj=0$)] initially
closely compensates the excess of the spatial delay
caused by the additional curvature, which
is $\dtsp = \rcol(0)/c - \rcol(\thj)\cos(\thj)/c
= \bla(0)\tcol(0) - \bla(\thj)\tcol(\thj)\cos(\thj)$. 
One can see that as long as $\gla(\thj) \gg 1$
and $\thj \la \thc$ the difference between them is
$\dtsp - \dttem \approx \dtsph$, where $\dtsph= r(0)[1 - \cos\thj]/c$
is the delay for the spherical curvature.
The exact value of
the temporal advance is slightly larger than
the additional spatial delay ($\dttem > \dtsp - \dtsph$),
so that photons emitted there are actually observed \emph{earlier}
than in the case of the spherical outflow.

\subsection{Outflow with the Lorentz factor independent of $\thj$}

In the cases with $\thj$-independent Lorentz factor (Figs.~\ref{f1}c
and \ref{f1}d) the lightcurves initially follow the slope $2 + \delta$
and then behave in a way which is easy to interpret.
For on-beam viewing ($\thobs \la \thc$) the decline of flux is faster
than $2 + \delta$, because the emissivity at a late time
(larger angles $\thj$) is weaker.
For off-beam viewing ($\thobs > \thc$) the drop of flux becomes slower,
because at the late time the bright core of the outflow becomes visible
for the observer.

Unlike in the case of $\thj$-dependent Lorentz factor,
the problem of non-simultanous turn-off does not arise in cases 2 and 4.

\section{Conclusions}
We have modeled the curvature effect in structured jet models. For
``on-beam'' viewing geometry, we find that the temporal decay slope,
$2+\delta$, as is derived for isotropic fireballs, is ubiquitous. The
deviations happen in the ``off-beam'' geometry, with the line of sight
pointing towards the Gaussian or power law wing of the structured jet.
This gives a shallower decay index than $2+\delta$.

Detailed analyses of {\em Swift} data indicate that in some
cases, the rapid decay slopes of X-ray early afterglow lightcurves
are somewhat shallower than $2+\delta$ (O'Brien et al. 2005). One
possible reason is that these reflect structured GRB jets viewed
``off-beam''.

\acknowledgments

This work was supported by a research grant at UNLV
and by 2P03D.004.24 (JD) and by NASA under grants NNG05GB67G, NNG05GH92G,
and NNG05GH91G (BZ).

\begin{thebibliography}{}
\bibitem[2005]{b05}Burrows, D. N., Romano, P., Falcone, A., Kobayashi,
S., Zhang, B. et al. 2005, Science, 309, 1833
\bibitem[2005]{c05}Chincarini, G. Moretti, A., Romano, P., Covino, S.,
Tagliaferri, G., et al. 2005, ApJ, submitted (astro-ph/0506453) 
\bibitem[2005]{fw05} Fan, Y. Z., \& Wei, D. M. 2005, \mnras, in press
(astro-ph/0506155)
\bibitem[2003]{gk03}Granot, J., \& Kumar, P. 2003, \apj, 591, 1086
\bibitem[2003]{kg03}Kumar, P., \& Granot, J. 2003, \apj, 591, 1075
\bibitem[2000]{kp2000}Kumar, P., \& Panaitescu, A. 2000, \apj, 541, L51
\bibitem[2005]{o05}O'Brien, P., Willingale, R., Goad, M. R., Page,
K. L. et al. 2005, to be submitted 
\bibitem[2005]{n05}Nousek, J. A., Kouveliotou, C., Grupe, D., Page, K.
et al. 2005, \apj, submitted (astro-ph/0508322)
\bibitem[2005]{pmg05} Panaitescu, A., M\'esz\'aros, P., Gehrels, N., 
  Burrows, D., \& Nousek, J.~2005, \mnras, submitted (astro-ph/0508340)
\bibitem[1994]{rm94}Rees, M. J., \& M\'esz\'aros, P. 1994, \apj, 430,
L93 
\bibitem[2002]{rlr02}Rossi, E., Lazzati, D. \& Rees, M. J. 2002,
\mnras, 332, 945 
\bibitem[2003]{s03}Salmonson, J. D. 2003, \apj, 592, 1002
\bibitem[2005]{t05}Tagliaferri, G., Goad, M., Chincarini, G.,
  Moretti, A., Campana, S., et al.~2005, Nature, 436, 985
\bibitem[2003]{wj03}Wei, D. M. \& Jin, Z. P. 2003, A\&A, 400, 415
\bibitem[2005]{yti05} Yamazaki, R., Toma, K., Ioka, K. \& Nakamura, T.
2005, \apj, submitted (astro-ph/0509159)
\bibitem[2005]{zdlm04}Zhang, B., Dai, X., Lloyd-Ronning, N. M. \&
M\'esz\'aros, P. 2004, \apj, 601, L119
\bibitem[2005]{zfd05}Zhang, B., Fan, Y. Z., Dyks, J., Kobayashi, S., 
  M\'esz\'aros, P., D. N. Burrows, J. A. Nousek, \& N. Gehrels, 2005,
\apj, submitted (astro-ph/0508321) 
\bibitem[2002]{zm02}Zhang, B. \& M\'esz\'aros, P. 2002, \apj, 571, 876
\end{thebibliography}
\end{document}